\documentclass[a4paper,11pt]{amsart}
\pdfoutput=1

\usepackage[utf8]{inputenc}
\usepackage[english]{babel}
\usepackage[T1]{fontenc}
\usepackage{microtype}
\usepackage[sb,rm,tt=false]{libertine}
\usepackage[lf,medium]{FiraSans}
\usepackage[T1]{fontenc}
\usepackage{textcomp}
\usepackage[varqu,varl]{zi4}
\usepackage{mathtools}
\usepackage[amsthm,libertine,vvarbb]{newtxmath}
\usepackage{thmtools}
\usepackage{color}
\usepackage{pgfplots}
\pgfplotsset{compat=1.18}
\usepackage{tikz}
\usetikzlibrary{arrows,shapes}
\usetikzlibrary{quantikz2,external}

\usepackage[numbers]{natbib}
\usepackage[hyphens]{url}

\usepackage[hidelinks]{hyperref}
\usepackage{titlesec}
\usepackage[ruled]{algorithm2e}
\usepackage[font=footnotesize]{subcaption}
\usepackage{enumitem}
\usepackage{mathdots}
\usepackage[nameinlink,noabbrev,capitalise]{cleveref}

\DisableLigatures[q]{shape = sc}

\newtheorem{theorem}{Theorem}

\theoremstyle{remark}
\newtheorem{remark}{Remark}

\Crefname{assumption}{Assumption}{Assumptions}
\theoremstyle{definition}
\newtheorem{definition}[theorem]{Definition}
\newtheorem{example}[remark]{Example}

\setlist[enumerate]{label=\arabic*.}

\definecolor{unia-mntf}{RGB}{0, 101, 97}
\definecolor{unia-purple}{RGB}{173, 0, 124}
\definecolor{unia-dark-blue}{RGB}{0, 135, 193}
\hypersetup{
    colorlinks=true,
    linkcolor=unia-purple,
    urlcolor=unia-purple,
    citecolor=unia-purple
}

\DeclareFontFamily{U}{matha}{\hyphenchar\font45}
\DeclareFontShape{U}{matha}{m}{n}{
	<-6> matha5 <6-7> matha6 <7-8> matha7
	<8-9> matha8 <9-10> matha9
	<10-12> matha10 <12-> matha12
}{}
\DeclareSymbolFont{matha}{U}{matha}{m}{n}

\DeclareFontFamily{U}{mathx}{\hyphenchar\font45}
\DeclareFontShape{U}{mathx}{m}{n}{
	<-6> mathx5 <6-7> mathx6 <7-8> mathx7
	<8-9> mathx8 <9-10> mathx9
	<10-12> mathx10 <12-> mathx12
}{}
\DeclareSymbolFont{mathx}{U}{mathx}{m}{n}

\DeclareMathDelimiter{\vvvert} {0}{matha}{"7E}{mathx}{"17}

\allowdisplaybreaks

\makeatletter
\def\th@plain{%
  \thm@headfont{\scshape}
  \thm@notefont{\normalfont}
  \itshape 
}
\def\th@definition{%
  \thm@headfont{\scshape}
  \thm@notefont{\normalfont}
  \normalfont 
}
\makeatother

\SetKwSty{textsc}

\makeatletter
\def\@setauthors{%
  \begingroup
  \def\thanks{\protect\thanks@warning}%
  \trivlist
  \raggedright\footnotesize \@topsep30\p@\relax
  \advance\@topsep by -\baselineskip
  \item\relax
  \author@andify\authors
  \def\\{\protect\linebreak}%
  \rmfamily\large\authors%
  \ifx\@empty\contribs
  \else
    ,\penalty-3 \space \@setcontribs
    \@closetoccontribs
  \fi
  \endtrivlist
  \endgroup
}
\def\@settitle{%
    \noindent
    \baselineskip14\p@\relax
    \LARGE
    \firabook\sffamily
    \@title
}
\def\@setaddresses{\par
  \nobreak \begingroup
  \def\author##1{\nobreak\addvspace\smallskipamount}%
  \def\\{\unskip, \ignorespaces}%
  \interlinepenalty\@M%
  \def\address##1##2{\begingroup%
    \par\addvspace\smallskipamount\indent%
    \@ifnotempty{##1}{(\ignorespaces##1\unskip) }%
    {\ignorespaces##2}\par\endgroup}%
  \def\curraddr##1##2{\begingroup%
    \@ifnotempty{##2}{\nobreak\indent\curraddrname
      \@ifnotempty{##1}{, \ignorespaces##1\unskip}\/:\space
      ##2\par}\endgroup}%
  \def\email##1##2{\begingroup
    \@ifnotempty{##2}{\nobreak\indent\emailaddrname
      \@ifnotempty{##1}{, \ignorespaces##1\unskip}\/:\space
      \ttfamily##2\par}\endgroup}%
  \def\urladdr##1##2{\begingroup
    \def~{\char`\~}%
    \@ifnotempty{##2}{\nobreak\indent\urladdrname
      \@ifnotempty{##1}{, \ignorespaces##1\unskip}\/:\space
      \ttfamily##2\par}\endgroup}%
\begin{flushleft}
    \addressfont\addresses
  \end{flushleft}
  \endgroup
}
\newcommand{\addressfont}{\normalfont\small}
\patchcmd{\abstract}{3pc}{0pt}{}{} 
\renewcommand{\@mkboth}[2]{} 
\patchcmd{\@maketitle}
  {\ifx\@empty\@dedicatory}
  {\@setaddresses\global\let\@setaddresses\relax\ifx\@empty\@dedicatory}
  {}{}
\makeatother

\titleformat{\section}
  {\sffamily\firabook\large} 
  {\thesection}
  {1em}
  {}
\titleformat{\subsection}
  {\sffamily\firabook\normalsize} 
  {\thesubsection}
  {1em}
  {}
\titlespacing*{\section} {0pt}{3.5ex plus 1ex minus .2ex}{2.3ex plus .2ex}
\titlespacing*{\subsection} {0pt}{3.25ex plus 1ex minus .2ex}{1.5ex plus .2ex}
\titleformat{\paragraph}[runin]{\normalfont\normalsize\itshape}{\theparagraph}{1em}{}[.]
\titlespacing*{\paragraph}{0pt}{3.25ex plus 1ex minus .2ex}{\the\fontdimen2\font}

\usepackage{listings}

\definecolor{brightred}{rgb}{1.00,0.00,0.00}
\definecolor{mediumgray}{rgb}{0.70,0.70,0.70}
\definecolor{olivegreen}{rgb}{0.49,0.56,0.16}
\definecolor{purple}{rgb}{0.68,0.00,0.49}
\definecolor{darkgreen}{rgb}{0.00,0.50,0.00}
\definecolor{steelblue}{rgb}{0.25,0.44,0.63}
\definecolor{darkteal}{rgb}{0.00,0.40,0.38}
\definecolor{brown}{rgb}{0.56,0.13,0.00}
\definecolor{brickred}{rgb}{0.73,0.13,0.13}
\definecolor{orange}{rgb}{0.74,0.48,0.00}
\definecolor{slateblue}{rgb}{0.02,0.16,0.49}
\definecolor{midblue}{rgb}{0.08,0.33,0.51}

\newcommand{\N}{\mathbb{N}}

\newcommand{\C}{\mathbb{C}}

\newcommand{\ie}{\eta}

\DeclareMathOperator{\Id}{Id}

\let\libraryfont\textsc
\newcommand{\Unitaria}{\libraryfont{Unitaria}}
\newcommand{\Tequila}{\libraryfont{Tequila}}

\SetKwSty{textsc}

\begin{document}
\title[Unitaria]{\Unitaria{}: Quantum Linear Algebra via Block Encodings}
\author[M.\ Deiml]{Matthias Deiml${}^{\dag}$}
\author[O.\ H\"uttenhofer]{Oliver H\"uttenhofer${}^{\ddagger}$}
\author[R.\ Mosco]{Ram Mosco${}^{\ddagger}$}
\author[J.\ S.\ Kottmann]{Jakob S.\ Kottmann${}^{\ddagger,\ast}$}
\author[D.\ Peterseim]{Daniel Peterseim${}^{\dag,\ast}$}
\address{${}^{\dag}$Institute of Mathematics, University of Augsburg, Germany}
\address{${}^{\ddagger}$Institute of Computer Science, University of Augsburg, Germany}
\address{${}^{\ast}$Centre for Advanced Analytics and Predictive Sciences, University of Augsburg, Germany}
\date{May 11, 2026}
\thanks{The work of M.\ Deiml and 
D.\ Peterseim is partially funded by the Deutsche Forschungsgemeinschaft (DFG, German Research Foundation) -- 571768116.}
\thanks{The work of O.\ H\"uttenhofer, R.\ Mosco and J.\ S.\ Kottmann is funded by the Federal Ministry of Research, Technology and Space (BMFTR) of Germany (Quantum Technologies/HoliQC2) -- 13N17231.}
\thanks{Main authors: M.\ Deiml, O.\ H\"uttenhofer}
\thanks{Corresponding authors: \texttt{jakob.kottmann@uni-a.de/daniel.peterseim@uni-a.de}}

\begin{abstract}
We introduce \Unitaria{}, a Python library that brings the simplicity of 
classical linear algebra toolkits such as \libraryfont{NumPy} and 
\libraryfont{SciPy} to the implementation of quantum algorithms based on block encodings, a general-purpose abstraction in which a matrix is embedded as a sub-block of a larger unitary operator. Their implementation has so far required deep knowledge of low-level circuit construction, which \Unitaria{} aims to eliminate. The library provides a composable, array-like interface through which users can define block encodings of matrices and vectors, combine them through standard operations such as addition, multiplication, tensor products, and the 
Quantum Singular Value Transformation, and extract the resulting quantum 
circuits automatically. A key feature is a matrix-arithmetic evaluation path in which every operation can be computed directly on encoded vectors and matrices without dependence on ancilla qubits or circuit simulation. This enables correctness verification and classical simulation that scale well beyond what state vector simulation permits and also allows resource estimation, including gate counts, qubit counts, and normalization constants, without executing any circuit. Together, these capabilities allow researchers to develop, verify, and analyze quantum linear algebra algorithms today, ahead of the availability of 
error-corrected hardware. \Unitaria{} is open source and available at 
\url{https://github.com/tequilahub/unitaria}.

\vspace{0.1cm}
\item[\hskip\labelsep\scshape{}Keywords.] block encodings, quantum linear 
algebra, quantum algorithms, Quantum Singular Value Transformation, quantum simulation, resource estimation, software library.

\vspace{0.1cm}
\item[\hskip\labelsep\scshape{}2020 Mathematics Subject Classification.] 15-04, 65-04, 74-04, 81-04.
\end{abstract}

\maketitle

\section{Introduction}
Block encodings have emerged as a central abstraction in quantum algorithm design, with applications spanning a wide range of scientific domains. From the mathematical perspective, they are an integral part of the solution of linear~\cite{MPS+25} and non-linear~\cite{DP24} systems of equations, optimization, and machine learning~\cite{RR23,GYC+24,ZZN+26}, as well as the treatment of partial differential equations~\cite{GHL25,SS25,DP25}, potentially involving multiple scales~\cite{HJZ24,BDP26}. Application areas include computational mechanics~\cite{LOC24}, fluid dynamics~\cite{LS25}, and quantum chemistry~\cite{MST+20,SBW+21,GWL+22,KMM23}. We can only list a limited number of examples here; the broader literature is vast.
This scientific breadth has driven growing demand for software tools that support the construction and analysis of block encodings. 
General-purpose frameworks include  \libraryfont{Qiskit}~\cite{Qiskit}, \libraryfont{Cirq}~\cite{Cirq}, \libraryfont{PennyLane}~\cite{Pennylane}, \libraryfont{Qrisp}~\cite{Qrisp}, and \libraryfont{Tequila}~\cite{Tequila}; simulators such as \libraryfont{Qulacs}~\cite{Qulacs} and \mbox{\libraryfont{CUDA-Q}~\cite{CudaQ}} target large-scale state vector simulation. 
These packages provide a strong foundation for the low-level design and simulation of gate-based quantum circuits. However, as the development of new quantum algorithms is progressing quickly, the demand for higher-level tools for their implementation is rapidly outgrowing the available libraries. 
This is particularly apparent for the construction of \emph{block encodings} -- a general-purpose abstraction in which a matrix is embedded as a subblock of a~larger unitary operator, enabling high-dimensional matrix computations on quantum computers. Manual implementation of these block encodings is tedious and error prone. Furthermore, while most existing quantum frameworks provide some surface-level functionality for block encodings, these interfaces often cover only rather rigid use cases.

The need for block encoding tools is reflected in the increase in scientific works, as well as in software projects related to this topic.
Simulation frameworks are increasingly adding support.
Notably, \libraryfont{Qrisp} recognized the need for operator-based construction of block encodings and their abstraction potential.~\cite{PZ26}
\libraryfont{Qualtran}~\cite{HKY+24} and \libraryfont{PennyLane} offer implementations of important building blocks but do not use the abstraction of block encodings to hide quantum-specific constructions. Instead, they add block encodings as an additional layer. Specialized tools such as \libraryfont{Fable}~\cite{CB22} enable the construction of concrete, intermediate-size matrices but do not implement the arithmetic operations needed to efficiently execute algorithms for high-dimensional applications.
To the best of our knowledge, mature implementations typically compile block encodings directly to quantum circuits, missing the opportunity to inspect, optimize, verify, and simulate higher-level structures.
The need for high-level optimization and the advantage of an operator-based syntax were, however, similarly observed by C.~Yuan~\cite{Yua25}, with the \libraryfont{Cobble} framework suggested therein, a~proof-of-concept implementation is available with further development on the way.

\Unitaria{} pursues this direction with a complete library of operations, a formalized subspace representation, and a matrix-arithmetic evaluation path that operates independently of circuit simulation.
We introduce \Unitaria{} to fill this gap in the quantum computing ecosystem. The aim of \Unitaria{} is to enable users to define and simulate block encodings with the simplicity of well-known linear algebra toolkits like \libraryfont{NumPy}, \libraryfont{SciPy}, or \libraryfont{MATLAB}. Objects are constructed through a familiar array-like interface, providing quantum circuits without requiring deep knowledge of low-level implementation details of quantum operations. We develop the library with three use cases in mind:
\begin{enumerate}
    \item We provide the infrastructure for defining block encodings on an abstract level, so that code written now can be used once error-corrected quantum computers become available. At the same time, this code can be efficiently verified on a matrix arithmetic level. This ensures that the algorithms are \emph{correct}.
    \item We enable researchers to calculate important performance metrics of their algorithms, namely gate count and the number of qubits, which are standard indicators for quantum circuits, as well as normalization constants that are specific to block encodings. Importantly, the matrix-arithmetic calculations and resource estimation require no circuit simulation and can thus be used even for previously out-of-reach quantum algorithms. This allows developers to assess whether an algorithm is \emph{efficient} ahead of hardware availability.
    \item Finally, we work towards defining an interoperable format for block-encoded matrices. This might allow the same block encodings written now in \Unitaria{} to be used across different programming languages, quantum computing frameworks, or, in the future, production-ready compilers, allowing code to be \emph{reusable}.
\end{enumerate}
Current quantum hardware does not yet permit the execution of circuits much beyond the simplest block encodings with meaningful accuracy~\cite{DP25}, which makes these three use cases particularly relevant today.

In this document, we describe the theoretical background of \Unitaria{} and outline its architecture. For the source code, installation instructions, and further documentation, visit \url{https://github.com/tequilahub/unitaria}.

\section{Block encodings}

In this section, we lay out the theoretical foundation that underlies \Unitaria{}, namely block encodings of matrices. The concept was introduced in~\cite{LC19}, while the name ``block encoding'' is due to~\cite{CGJ19}. Originally, the intention was to encode matrices for performing Hamiltonian simulation and, by extension, solving linear systems of equations using Quantum Signal Processing and the Quantum Singular Value Transformation~\cite{LC19,LC17,GSLW19,MRTC21}, but their usefulness now far exceeds these specific techniques and applications.

\subsection{Definition} \label{sec:block_encodings}
Unitary matrices can be encoded naturally on quantum computers. Namely, for any unitary matrix~$U$, there is a quantum circuit that transforms the quantum state by the action of $U$, or at least approximates that action.
The same, however, is not possible for an arbitrary non-unitary matrix~$A$. To overcome this, the idea of block encodings is to find a circuit such that its unitary action~$U$ contains $A$ as a block, i.e.
\[
U = \begin{bmatrix}
A & \ast \\ \ast & \ast
\end{bmatrix}.
\]
Still, such a construction requires $\|A\|_2 \le 1$, where $\|\bullet\|_2$ specifies the spectral norm. This approach can be extended to all matrices by allowing for a normalization factor $\gamma \ge \|A\|_2$ giving
\begin{equation} \label{eq:block-encoding}
U = \begin{bmatrix}
    A / \gamma & \ast \\ \ast & \ast
\end{bmatrix}.
\end{equation}
This leads to the now common definition of an ``$(\alpha, a, \varepsilon)$-block-encoding''~\cite[Definition~43]{GSLW19} as an approximation
\[
\|A - \alpha(\bra{0}_a \otimes \Id) U (\ket{0}_a \otimes \Id)\|_2 \le \varepsilon.
\]
This definition has a few caveats. It only allows for square matrices, restricts the position of $A$ in the upper left block of $U$ as in~\eqref{eq:block-encoding}, and by the parameter $\varepsilon$ mixes the questions of encoding and approximation. We thus use the following, slightly modified definition:

\begin{definition}[Block encoding]
Let $m \in \N$ be a number of qubits and $A \in \C^{N_\mathrm{out} \times N_\mathrm{in}}$ be a matrix with dimensions $N_\mathrm{in}, N_\mathrm{out} \le 2^m$. Consider a decomposition
\[
A_{k\ell} = \gamma\bra{j_k^\mathrm{out}}U\ket{j_\ell^\mathrm{in}}.
\]
Here $0 \le j_0^{\ast} < \dots < j^{\ast}_{N_\ast - 1} < 2^m$ for $\ast \in \{\mathrm{in}, \mathrm{out}\}$ are ordered sequences of indices, and $\gamma \ge \|A\|_2$ is a positive real number called the \emph{normalization}. Then, a \emph{block encoding} of $A$ is a data structure containing the normalization, e.g., as a float, a quantum circuit corresponding to $U$, and the sequences $j_0^\ast, \dots, j_{N_\ast-1}^\ast$, potentially in a compressed format; see \cref{sec:subspaces}.
\end{definition}

\begin{example}[Matrix with all ones]
Consider the $2\times 2$ matrix consisting of only ones
\[
A = \begin{bmatrix}
    1 & 1 \\
    1 & 1
\end{bmatrix},
\]
which clearly is not unitary. We can still construct a block encoding of $A$. This matrix can be seen as the sum
\[
A = {\Id} + X
\]
of the identity and the Pauli $X$ matrix. Block encodings of such sums can be obtained through the Linear Combination of Unitaries (LCU) technique, which gives the circuit
\[\setlength\arraycolsep{4pt}
U_A = \begin{quantikz}[column sep={0.3cm}, row sep={0.6cm,between origins}]
        && \targ{} & & \\
        &\gate{H} & \ctrl{-1} & \gate{H} &
    \end{quantikz} = \frac12 \begin{bmatrix}
    \phantom{-}1 & \phantom{-}1 & \phantom{-}1 & -1 \\
    \phantom{-}1 & \phantom{-}1 & -1 & \phantom{-}1 \\
    \phantom{-}1 & -1 & \phantom{-}1 & \phantom{-}1 \\
    -1 & \phantom{-}1 & \phantom{-}1 & \phantom{-}1
\end{bmatrix}.
\]
Clearly, $A$ is the upper-left block of $U$ up to a factor of $2$. This means that if we choose $\gamma = 2$ and $j_k^\ast = \ket{k}$ for $k \in \{0, 1\}$ and $\ast \in \{\mathrm{in}, \mathrm{out}\}$, then we obtain a block encoding of $A$.
\end{example}

A block encoding thus considers the part of the unitary $U$ for which the input vector belongs to the $N_\mathrm{in}$-dimensional subspace of $\C^{2^m}$ spanned by $\ket{\smash{j_0^\mathrm{in}}}, \dots \ket{\smash{j_{N_\mathrm{in}}^\mathrm{in}}}$ , and the output vector belongs to the $N_\mathrm{out}$-dimensional subspace spanned by $\ket{\smash{j_0^\mathrm{out}}}, \dots \ket{\smash{j_{N_\mathrm{out}}^\mathrm{out}}}$.
If $N_\mathrm{in} = 1$, meaning $A$  consists of a single column, we can interpret $A$ as a vector. By this logic -- equating column matrices with vectors -- we extend our definition to any vector. This coincides with what is known as ``state preparation'' circuits in quantum computing.

Instead of the index based decomposition above, we may alternatively define projectors
\[
\Pi_\ast \coloneqq \sum_{k = 0}^{N_\ast - 1} \ket{k}\bra{\smash{j_k^\ast}}, \quad *\in\left\{\mathrm{in}, \mathrm{out}\right\}
\]
and write $A = \gamma\Pi_\mathrm{out} U \Pi_\mathrm{in}^\dag$. This gives rise to a slightly more general definition of block encodings considered in \cite{DP24,DP25}. However, it is unclear in what format these projections should be given, making it harder to judge computational complexity. In \Unitaria{}, we stick to the wording of \emph{subspaces} but note that we only allow subspaces that correspond to computational basis states; in other words, the index sets defined above. In this way, we can define a concrete format, explained below, in which subspaces are stored.

Note that -- at best -- the number of gates in a quantum circuit can be linear in $m$ and logarithmic in the dimension $2^m$. The same is true for the index sequences if a good format is used. Thus, block encodings allow us to describe matrices in a number of bits that is \emph{polylogarithmic} in the dimension. This is also true for some classical tensor formats, but computations using other classical encodings, such as sparse matrices, typically require memory that is at least linear in the dimension of the matrix.

This exponential memory reduction is already interesting, but what makes block encodings useful is the possibility of also combining and evaluating them in polylogarithmic time.

\subsection{Construction of basic operations}

We introduce some basic operations on block encodings that are implemented in \Unitaria{}. The inputs to these operations are block encodings of one or more matrices $A, B, C, \dots$. From this, we define programmatic ways to construct block encodings of, e.g., $AB$, $A + B$, or $A \otimes B$.

\paragraph{Adjoint}
Most of the constructions involve technical details that may not be interesting to the reader; for details, we refer to the implementation, \cite[Section~4]{GSLW19}, and \cite[Appendix~B]{DP25}. The adjoint of a matrix~$A^\dag$, however, is quite simple. Let us consider a block encoding
\[
A_{k\ell} = \gamma\bra{j_k^\mathrm{out}}U\ket{j_\ell^\mathrm{in}}.
\]
Then, for the adjoint, it holds that
\[
A_{k\ell}^\dag = \overline{A_{\ell k}} =
\gamma\overline{\bra{j_\ell^\mathrm{out}}U\ket{j_k^\mathrm{in}}} =
\gamma\bra{j_k^\mathrm{in}}U^\dag\ket{j_\ell^\mathrm{out}}.
\]
This defines a block encoding for $A^\dag$, since a circuit for the adjoint $U^\dag$ is obtained simply by reversing the circuit for $U$ and inverting each gate.

\paragraph{Multiplication}
The most useful, but perhaps the most complex operation to implement is matrix-matrix (or, equivalently, matrix-vector) multiplication. On a conceptual level, the quantum circuit for a~product of block encodings with circuits $U_A$ and $U_B$ is simply given by the sequential application of these circuits~$U_AU_B$. However, this only works if the index sets or subspaces of the block encodings align. Always choosing subspaces as $\ket{0}, \dots, \ket{N-1}$ would resolve this issue, but this constraint would lead to inefficiencies in the resulting circuits. Thus, \Unitaria{} internally applies permutations between $U_A$ and $U_B$ to convert between the different index sets. Finally, in the general case, one needs to check whether the intermediate state between $U_A$ and $U_B$ is in the correct subspace, but often this check can be skipped.

\paragraph{Tensor products}
Tensor products are again a very natural construction in quantum computing, as the parallel execution of two circuits $U_A$ and $U_B$ on separate qubits gives rise to the unitary~$U_A \otimes U_B$. A block encoding of the tensor product of two matrices is obtained by choosing the correct normalizations and index sets.

\paragraph{Block diagonals}
Block diagonal matrices are not necessarily important per se, but they are used in the construction of many other operations, most importantly addition. Given circuits $U_A$ and~$U_B$ the corresponding block diagonal unitary is implemented using the controlled applications
\[
\begin{bmatrix}
    U_A & 0 \\ 0 & U_B
\end{bmatrix} = (X \otimes \Id)CU_A(X \otimes \Id)CU_B.
\]
Again, this gives rise to the desired block encoding, as long as the normalizations of the encodings of~$A$ and~$B$ match.

\paragraph{Addition}
The operations above are indeed already enough to implement addition by the decomposition
\begin{equation} \label{eq:add}
A + B =
\left(\begin{bmatrix}
    \sqrt{\gamma_A} & \sqrt{\gamma_B}
\end{bmatrix} \otimes \Id\right)
\begin{bmatrix}
    A/\gamma_A & 0 \\ 0 & B/\gamma_B
\end{bmatrix}
\left(
\begin{bmatrix}
    \sqrt{\gamma_A} \\ \sqrt{\gamma_B}
\end{bmatrix} \otimes \Id
\right),
\end{equation}
which is equivalent to the Linear Combination of Unitaries (LCU)~\cite{BCC+15} technique.
A block encoding can thus be found using the previously defined block diagonal, tensor, and multiplication operations.

\subsection{Quantum Singular Value Transformation}

Quantum Singular Value Transformation~(QSVT)~\cite{GSLW19,MRTC21} and its predecessor, Quantum Signal Processing~(QSP)~\cite{LC17,LC19}, were the original motivation for block encodings and are surely among the most important building blocks for quantum computing, encompassing important procedures such as Grover search~\cite{Gro98}. Concretely, QSVT allows us to compute matrix polynomials of any block encoded matrix with almost optimal information efficiency. This is done by interleaving applications of the block encoding unitary with rotations around the input and output subspaces.

However, QSVT circuits can be tedious to implement manually. The computation of rotation angles is complicated and requires external tools such as \libraryfont{pyqsp}~\cite{MRTC21} or \libraryfont{QSPPack}~\cite{DMWL21}, which might have different conventions on the axis of rotation. Handling high-degree polynomials requires care due to the loss of numerical accuracy, and errors are often hard to debug. For this reason, a~unified framework like \Unitaria{} can help, offering an interface for defining block encodings and alleviating the user from having to deal with rotation angles.

\subsection{The importance of normalization}
To assess the quality of block encodings, it is important to understand the interplay of time complexity and success probability. We quantify the latter using the \emph{information efficiency}
\[
\ie(A) \coloneqq \|A\|_2 / \gamma \le 1. 
\]
If, for example, we use na\"ive projective measurement with a block encoding, the success probability -- the probability of obtaining a state that is $\ket{j_k^\mathrm{out}}$ for some $k$ -- is upper bounded by the square of the information efficiency~$\ie(A)^2$. By using more sophisticated methods such as \emph{amplitude amplification} \cite{BHMT02}, the complexity of obtaining a state in the selected subspace is only linear in $\ie(A)^{-1}$. Still, low information efficiency can be detrimental to the efficiency of an algorithm.

This effect can also be seen when attempting to measure the Euclidean norm~$\|v\|_2$ of a block-encoded vector~$v$. This will require a number of queries to the block encoding proportional to $\varepsilon^{-1} \ie(v)^{-1} \log(\delta^{-1})$, where $\varepsilon > 0$ is the relative error tolerance, and $0 < \delta < 1$ is the maximum allowed failure probability.

The takeaway here is that one should always try to keep the normalization factor~$\gamma$ as low as possible. At the same time, it is important to note that this is not about choosing an appropriate~$\gamma$. The difficulty is rather in finding the correct circuit $U$. The normalization is then the ratio between the actual quantum implementation and the target matrix and cannot be chosen freely.

\Unitaria{} cannot yet optimize block encodings to improve the normalization by itself. However, it simplifies the computation of both the normalization and the gate count, which together allow estimating the runtime of a quantum algorithm without actually executing it. This makes the testing of complex quantum algorithms possible, even when the required number of qubits exceeds the capabilities of state vector simulators.

\section{Implementation}

In the following, we provide a brief high-level summary of \Unitaria{}'s design.
This is not meant to replace \Unitaria{}'s developer documentation at \url{https://tequilahub.github.io/unitaria/docs}, which describes all available classes and functions, but rather to give an overview of \Unitaria{}'s software architecture and make it easier to get started with the documentation and code examples.

\subsection{Subspaces} \label{sec:subspaces}

Subspaces are encoded in \Unitaria{} by the \texttt{Subspace} class.
It stores index subsets, i.e.\ the indices $0 \le j_0^* < \dots < j_{N_*-1}^* < 2^m$ for $* \in \{\mathrm{in}, \mathrm{out}\}$ of a block encoding as described in \cref{sec:block_encodings}, in a compressed format.
Specifically, the space spanned by $\ket{\smash{j_0^*}}, \dots, \ket{\smash{j_{N_*-1}^*}}$ is decomposed into a~tensor product, corresponding to a partition of the $m$ qubits into distinct subsets.

Objects of the \texttt{Subspace} class store a list of these tensor factors, ordered from least to most significant.
Each tensor factor has one of two types. The first is a \texttt{ZeroQubitSubspace}, which indicates that this qubit must be in the state $\ket{0}$ for all vectors in the subspace.
The second type of factor is a~\texttt{ControlledSubspace}, where the most significant qubit controls in which of the two smaller subspaces the other qubits must be.
This recursive structure means that these two types of tensor factors are powerful enough to represent all subspaces required for the block encodings in \Unitaria{}, while being simple enough that circuits can be automatically generated to check whether a state lies in the subspace.
There is no need for a separate factor type to indicate the full space of a single qubit, since this can be achieved by a \texttt{ControlledSubspace} where the two smaller subspaces are the zero-qubit space $\C^{2^0}$. The latter is in turn represented by a \texttt{Subspace} with an empty list of tensor factors.

The preferred way to define subspaces is using an operator based syntax, which allows constructing a \texttt{ControlledSubspace} through the operator~\texttt{|}, and tensor products through the operator~\texttt{\&}. This is complemented by a constructor given a string of the characters \texttt{0} and \texttt{\#}, representing $\ket{0}$ and $\C^2$ respectively.
For example, consider the subspace spanned by the basis vectors $\{\ket{0000}, \ket{1000}, \ket{1010}\}$.
This subspace is spanned by the states of the form
\[ \left(\ket{j} \otimes \begin{cases}\ket{00} &\text{if $j = 0$}\\ \ket{0} \otimes \ket{k} &\text{if $j = 1$}\end{cases}\right) \otimes \ket{0} \quad\text{for } j,k \in \{0, 1\}\]
and is implemented by the following code:
\begin{lstlisting}[language=Python]
(ut.Subspace("00") | ut.Subspace("0#")) & ut.Subspace("0")
\end{lstlisting}
In the internal format, which we recall uses the reverse order for the tensor factors, the constructed object is stored as
\begin{lstlisting}[language=Python]
Subspace([
    ZeroQubitSubspace(),
    ControlledSubspace(
        Subspace([ZeroQubitSubspace(), ZeroQubitSubspace()]),
        Subspace([
            ControlledSubspace(Subspace([]), Subspace([])),
            ZeroQubitSubspace(),
        ]),
    )
])
\end{lstlisting}

Note that not all subspaces can be represented -- as a simple example, you cannot define the one-qubit subspace spanned by the $\ket{1}$ state.
However, this is not necessary for the block encodings used in \Unitaria{} and this restriction allows for more efficient circuit implementations. Indeed, users do not need to construct concrete subspaces and can mostly rely on the convenience function \texttt{Subspace.from\_dim} to construct a Subspace of a given dimension. Further, if a sub-block of an encoded matrix is required, users can resort to index notation of the form \texttt{U[:x,:y]}, which automatically constructs the correct subspaces. Specifying more complex subspaces should only be needed for defining new low-level block encodings or for manually optimizing the allocation of qubits.

\subsection{Nodes}

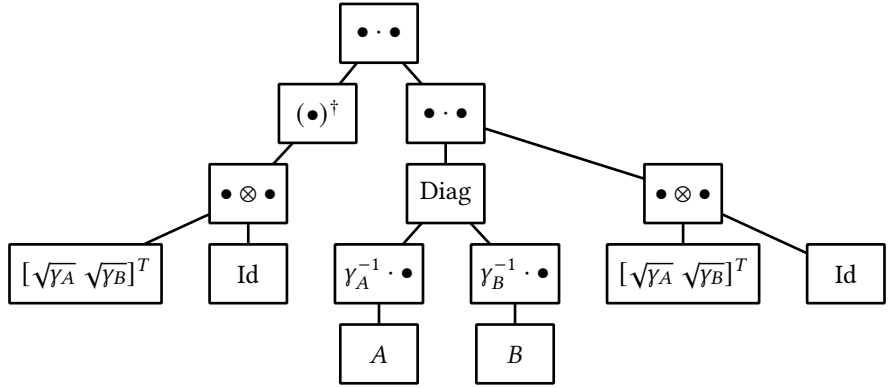
\begin{figure}
\begin{center}
\begin{tikzpicture}[>=latex,line join=bevel,]
  \pgfsetlinewidth{1bp}
\pgfsetcolor{black}
  \draw [] (130.2bp,119.89bp) .. controls (128.14bp,117.39bp) and (125.93bp,114.7bp)  .. (123.87bp,112.19bp);
  \draw [] (147.53bp,119.89bp) .. controls (149.76bp,117.39bp) and (152.16bp,114.7bp)  .. (154.4bp,112.19bp);
  \draw [] (106.11bp,89.891bp) .. controls (103.79bp,87.385bp) and (101.29bp,84.698bp)  .. (98.96bp,82.188bp);
  \draw [] (74.734bp,63.222bp) .. controls (67.541bp,59.92bp) and (58.74bp,55.881bp)  .. (50.754bp,52.215bp);
  \draw [] (89.5bp,59.891bp) .. controls (89.5bp,57.385bp) and (89.5bp,54.698bp)  .. (89.5bp,52.188bp);
  \draw [] (163.5bp,89.891bp) .. controls (163.5bp,87.385bp) and (163.5bp,84.698bp)  .. (163.5bp,82.188bp);
  \draw [] (178.2bp,95.376bp) .. controls (194.69bp,90.188bp) and (221.28bp,81.821bp)  .. (237.78bp,76.63bp);
  \draw [] (154.47bp,59.891bp) .. controls (152.24bp,57.385bp) and (149.84bp,54.698bp)  .. (147.6bp,52.188bp);
  \draw [] (172.89bp,59.891bp) .. controls (175.21bp,57.385bp) and (177.71bp,54.698bp)  .. (180.04bp,52.188bp);
  \draw [] (138.5bp,29.891bp) .. controls (138.5bp,27.385bp) and (138.5bp,24.698bp)  .. (138.5bp,22.188bp);
  \draw [] (189.5bp,29.891bp) .. controls (189.5bp,27.385bp) and (189.5bp,24.698bp)  .. (189.5bp,22.188bp);
  \draw [] (252.5bp,59.891bp) .. controls (252.5bp,57.385bp) and (252.5bp,54.698bp)  .. (252.5bp,52.188bp);
  \draw [] (267.27bp,63.222bp) .. controls (276.86bp,58.82bp) and (289.31bp,53.105bp)  .. (298.87bp,48.716bp);
\begin{scope}
  \definecolor{strokecol}{rgb}{0.0,0.0,0.0};
  \pgfsetstrokecolor{strokecol}
  \draw (153.0bp,142.0bp) -- (124.0bp,142.0bp) -- (124.0bp,120.0bp) -- (153.0bp,120.0bp) -- cycle;
  \draw (138.5bp,131.0bp) node {\small $\bullet \cdot \bullet$};
\end{scope}
\begin{scope}
  \definecolor{strokecol}{rgb}{0.0,0.0,0.0};
  \pgfsetstrokecolor{strokecol}
  \draw (130.0bp,112.0bp) -- (101.0bp,112.0bp) -- (101.0bp,90.0bp) -- (130.0bp,90.0bp) -- cycle;
  \draw (115.5bp,101.0bp) node {\small $(\bullet)^\dagger$};
\end{scope}
\begin{scope}
  \definecolor{strokecol}{rgb}{0.0,0.0,0.0};
  \pgfsetstrokecolor{strokecol}
  \draw (178.0bp,112.0bp) -- (149.0bp,112.0bp) -- (149.0bp,90.0bp) -- (178.0bp,90.0bp) -- cycle;
  \draw (163.5bp,101.0bp) node {\small $\bullet \cdot \bullet$};
\end{scope}
\begin{scope}
  \definecolor{strokecol}{rgb}{0.0,0.0,0.0};
  \pgfsetstrokecolor{strokecol}
  \draw (104.0bp,82.0bp) -- (75.0bp,82.0bp) -- (75.0bp,60.0bp) -- (104.0bp,60.0bp) -- cycle;
  \draw (89.5bp,71.0bp) node {\small $\bullet \otimes \bullet$};
\end{scope}
\begin{scope}
  \definecolor{strokecol}{rgb}{0.0,0.0,0.0};
  \pgfsetstrokecolor{strokecol}
  \draw (57.0bp,52.0bp) -- (0.0bp,52.0bp) -- (0.0bp,30.0bp) -- (57.0bp,30.0bp) -- cycle;
  \draw (28.5bp,41.0bp) node {\small $[\sqrt{\gamma_A}\; \sqrt{\gamma_B}]^T$};
\end{scope}
\begin{scope}
  \definecolor{strokecol}{rgb}{0.0,0.0,0.0};
  \pgfsetstrokecolor{strokecol}
  \draw (104.0bp,52.0bp) -- (75.0bp,52.0bp) -- (75.0bp,30.0bp) -- (104.0bp,30.0bp) -- cycle;
  \draw (89.5bp,41.0bp) node {\small $\operatorname{\Id}$};
\end{scope}
\begin{scope}
  \definecolor{strokecol}{rgb}{0.0,0.0,0.0};
  \pgfsetstrokecolor{strokecol}
  \draw (178.0bp,82.0bp) -- (149.0bp,82.0bp) -- (149.0bp,60.0bp) -- (178.0bp,60.0bp) -- cycle;
  \draw (163.5bp,71.0bp) node {\small $\operatorname{Diag}$};
\end{scope}
\begin{scope}
  \definecolor{strokecol}{rgb}{0.0,0.0,0.0};
  \pgfsetstrokecolor{strokecol}
  \draw (267.0bp,82.0bp) -- (238.0bp,82.0bp) -- (238.0bp,60.0bp) -- (267.0bp,60.0bp) -- cycle;
  \draw (252.5bp,71.0bp) node {\small $\bullet \otimes \bullet$};
\end{scope}
\begin{scope}
  \definecolor{strokecol}{rgb}{0.0,0.0,0.0};
  \pgfsetstrokecolor{strokecol}
  \draw (155.0bp,52.0bp) -- (122.0bp,52.0bp) -- (122.0bp,30.0bp) -- (155.0bp,30.0bp) -- cycle;
  \draw (138.5bp,41.0bp) node {\small $\gamma_A^{-1} \cdot \bullet$};
\end{scope}
\begin{scope}
  \definecolor{strokecol}{rgb}{0.0,0.0,0.0};
  \pgfsetstrokecolor{strokecol}
  \draw (206.0bp,52.0bp) -- (173.0bp,52.0bp) -- (173.0bp,30.0bp) -- (206.0bp,30.0bp) -- cycle;
  \draw (189.5bp,41.0bp) node {\small $\gamma_B^{-1} \cdot \bullet$};
\end{scope}
\begin{scope}
  \definecolor{strokecol}{rgb}{0.0,0.0,0.0};
  \pgfsetstrokecolor{strokecol}
  \draw (153.0bp,22.0bp) -- (124.0bp,22.0bp) -- (124.0bp,0.0bp) -- (153.0bp,0.0bp) -- cycle;
  \draw (138.5bp,11.0bp) node {\small $A$};
\end{scope}
\begin{scope}
  \definecolor{strokecol}{rgb}{0.0,0.0,0.0};
  \pgfsetstrokecolor{strokecol}
  \draw (204.0bp,22.0bp) -- (175.0bp,22.0bp) -- (175.0bp,0.0bp) -- (204.0bp,0.0bp) -- cycle;
  \draw (189.5bp,11.0bp) node {\small $B$};
\end{scope}
\begin{scope}
  \definecolor{strokecol}{rgb}{0.0,0.0,0.0};
  \pgfsetstrokecolor{strokecol}
  \draw (281.0bp,52.0bp) -- (224.0bp,52.0bp) -- (224.0bp,30.0bp) -- (281.0bp,30.0bp) -- cycle;
  \draw (252.5bp,41.0bp) node {\small $[\sqrt{\gamma_A}\; \sqrt{\gamma_B}]^T$};
\end{scope}
\begin{scope}
  \definecolor{strokecol}{rgb}{0.0,0.0,0.0};
  \pgfsetstrokecolor{strokecol}
  \draw (328.0bp,52.0bp) -- (299.0bp,52.0bp) -- (299.0bp,30.0bp) -- (328.0bp,30.0bp) -- cycle;
  \draw (313.5bp,41.0bp) node {\small $\operatorname{\Id}$};
\end{scope}
\end{tikzpicture}
\end{center}
\vspace{-1cm}
\caption{\label{fig:add}Computational graph for the implementation of addition $A + B$ \eqref{eq:add}.}
\end{figure}

\Unitaria{} does not execute computations immediately, but rather collects operations into a computational graph. The computational graph for the decomposition of the addition~\eqref{eq:add} is, for example, displayed in \cref{fig:add}.

Each operation supported by \Unitaria{} is thus defined as a subclass of the abstract \texttt{Node} class.
Implementing a subclass of \texttt{Node} requires functions for applying the operation and its adjoint on a vector or matrix and for building the corresponding quantum circuit.
Furthermore, subclasses must provide methods for querying the input and output subspaces, the normalization, and the number of ancillae used by the operation.
The \texttt{Node} class provides caching for most of the functions, so expensive operations like circuit construction do not need to be run multiple times.
Furthermore, various convenience functions are provided, for example, to check if a node is a~vector or to draw a diagram of the node and its children.

\Unitaria{} provides implementations for many types of nodes, from basic matrix operations such as addition, multiplication, or scaling, to more complex operations like integer operations, the Fourier Transform and the QSVT.
Often, it is useful to define nodes in terms of other nodes instead of implementing them explicitly, which is made possible by the \texttt{ProxyNode} class.
For example, fixed-point amplification, singular value amplification, and the Moore--Penrose inverse of a node can all be implemented by applying the QSVT with different polynomials, which is made simple by defining them as subclasses of \texttt{ProxyNode}.
The most commonly used operations should be covered by the nodes provided in \Unitaria{}, but for more specialized operations, users can implement custom nodes by subclassing \texttt{Node} or \texttt{ProxyNode}.

To allow for more concise and readable code, many operators of the \texttt{Node} class are overloaded, this includes \texttt{+}, \texttt{-}, \texttt{@}, and \texttt{*} for addition, subtraction, matrix-matrix multiplication, and scalar multiplication. The \texttt{\&} operator represents a tensor product, while \texttt{|} can be used to construct block diagonal matrices. Indexing of the form \texttt{A[a:b:c, x:y:z]} is also possible, and is internally implemented by constructing the corresponding projection matrix. The adjoint $(\bullet)^\dag$, i.e.\ transposed complex conjugate, is available through \texttt{A.adjoint()}.

\subsection{Matrix arithmetic evaluation}

Working with high-level block encodings instead of building circuits from gates is not only faster and less error prone, but also means that \Unitaria{} has semantic information that would be very difficult to extract from gate-based circuits.
This allows us to classically simulate most types of nodes by implementing the operation directly instead of executing each gate individually, and also to avoid ancilla qubits, which would incur exponential overhead when simulating circuits classically.
For example, integer addition, which adds a constant to the basis state, can be simulated on a statevector \texttt{input} using the following line of \libraryfont{NumPy} code:
\begin{lstlisting}[language=Python]
result = np.roll(input, self.constant)
\end{lstlisting}
This is significantly faster than explicitly simulating each gate in the corresponding circuit.
Even though the main purpose of \Unitaria{} is to map operations to quantum circuits, being able to efficiently simulate those operations classically can be invaluable for prototyping and testing.
The built-in verification function \texttt{ut.verify} ensures that the result of the direct implementation matches the behavior of the circuit by comparing the results on the standard basis.

\subsection{Backends and integration}

\Unitaria{} uses the quantum algorithm framework \Tequila{}~\cite{Tequila} to represent and run quantum circuits.
Through \Tequila{}, we gain access to multiple backends, including interfaces to quantum processors through \libraryfont{AQT}~\cite{AQT} or \libraryfont{Qiskit}~\cite{Qiskit}, as well as simulators like \libraryfont{Qulacs}~\cite{Qulacs}, which is used as the default simulator, or \libraryfont{CUDA-Q}~\cite{CudaQ}.
The software architecture is designed to allow support for other backends (such as \libraryfont{Qrisp} or \libraryfont{PennyLane}) in the future. Only core operations build circuits directly, allowing easy integration of other circuit backends.
For classical calculations, \Unitaria{} uses \libraryfont{NumPy}~\cite{Numpy} and \libraryfont{SciPy}~\cite{SciPy}.

\section{Examples}

In this section, we demonstrate the usage of \Unitaria{} through code examples. We start with the basic usage of \texttt{Node} objects and then turn to two practical examples, namely the solution of a~partial differential equation and the implementation of convolutions.

\subsection{A simple increment circuit}

As a basic usage demonstration, let us look at the \texttt{Increment} node, which implements the (unitary) linear map
\[
\ket{k}_n \mapsto \ket{k + 1 \operatorname{mod} 2^n}_n
\]
where $n \in \N$ is the number of qubits. We can extract the basic components of the block encoding, the circuit, subspaces, and normalization, using the methods of the object.

\noindent\begin{minipage}{\linewidth}
\begin{lstlisting}[language=Python]
>>> import unitaria as ut
>>> inc = ut.Increment(bits=2)
>>> inc.circuit()
Circuit(_tq_circuit=circuit: 
X(target=(1,), control=(0,))
X(target=(0,))
, n_qubits=2)
>>> inc.subspace_in.enumerate_basis()
array([0, 1, 2, 3], dtype=int32)
>>> inc.normalization
1
\end{lstlisting}
\end{minipage}

We can simulate the circuit using \texttt{simulate}, and use the matrix-arithmetic implementation with the methods \texttt{compute} and \texttt{toarray}. Their output should be almost identical. These methods also take care of projecting the simulated state vector to the correct subspace.

\noindent\begin{minipage}{\linewidth}
\begin{lstlisting}[language=Python]
>>> import numpy as np
>>> inc.simulate(np.array([0, 1, 0, 0], dtype=complex))
array([0.+0.j, 0.+0.j, 1.+0.j, 0.+0.j])
>>> inc.compute(np.array([0, 1, 0, 0], dtype=complex))
array([0.+0.j, 0.+0.j, 1.+0.j, 0.+0.j])
>>> inc.toarray()  # same output as inc.simulate()
array([[0.+0.j, 0.+0.j, 0.+0.j, 1.+0.j],
       [1.+0.j, 0.+0.j, 0.+0.j, 0.+0.j],
       [0.+0.j, 1.+0.j, 0.+0.j, 0.+0.j],
       [0.+0.j, 0.+0.j, 1.+0.j, 0.+0.j]])
\end{lstlisting}
\end{minipage}

The real power of \Unitaria{}, however, lies in combining block encodings, for which we will look at two more real-world examples.

\subsection{Solving a partial differential equation using the Finite Element Method}

As a further example, we illustrate the construction and inversion of the 1D discrete Laplace operator.
On the interval $D = [0, 1]$, the Laplace equation with homogeneous boundary condition is given by
\begin{align*}
-\Delta u &= f && \text{in }D \\
u &= 0 && \text{on }\partial D
\end{align*}
We solve for the function $u \in H^1_0(D)$ given the right-hand side $f \in L^2(D)$.
After discretization with standard P1 elements, we obtain a linear system of the form $Ax = b$ with
\[
A = 2^N\begin{bmatrix}
2 & -1 \\
-1 & \ddots & \ddots \\
& \ddots&& -1 \\
&& -1 & 2
\end{bmatrix}.
\]
We choose a right-hand side encoding $f \equiv 1$ and measure a quantity approximating the $L^2$ norm of the solution~$u$.

\noindent\begin{minipage}{\linewidth}
\begin{lstlisting}[language=Python]
import numpy as np
import unitaria as ut

N = 3
dofs = 2**N - 1

I = ut.Identity(dim=2**N)
X = ut.Increment(bits=N)

A = 2**N * (2 * I - X.adjoint() - X)[:-1, :-1]

vec2d = ut.ConstantVector(1 / 2 * np.ones(2))
b = (vec2d & vec2d & vec2d)[:-1]

condition = np.linalg.cond(A.toarray(), p=2)
A_inv = ut.Pseudoinverse(A, condition=condition, tolerance=0.01)

solution = A_inv @ b

qoi = solution.simulate_norm() * 2 ** (-N / 2)
\end{lstlisting}
\end{minipage}

A more sophisticated approach would be using the preconditioner from \cite{DP25}.
While the construction for the right-hand side $b$ is already in a form that would scale logarithmically with the number of degrees of freedom, an actual algorithm should not numerically attempt to compute the condition number of $A$. For this purpose, we plan to integrate the adaptive solvers from \cite{DP26}.
For a more detailed explanation of this code, see \href{https://tequilahub.github.io/unitaria/tutorials/partial_differential_equation/partial_differential_equation.html}{the tutorial notebook}.

\subsection{Convolution}
A convolution is defined as an integral operator with a kernel $K$
\begin{equation*}
    (K \ast f)(x) := \int K(x-y) f(y) \operatorname{d}y
\end{equation*}
which, when discretized, takes the form of a matrix-vector operation. 
A typical example are Gaussian convolutions with kernels $K(x - y) = e^{-a |x - y|^2}$ that have a broad range of applications (smoothing, filtering, approximating operators) within various fields, and are interesting due to their separability which simplifies high-dimensional implementations~\cite{BM05}.
While this approach originates from classical computing, it naturally extends to quantum computing where applying multiple single dimensional functions to separate registers results in their tensor product.

The matrix representation of a convolution is a Toeplitz matrix, e.g. on two qubits and for the kernel $K(x - y) = e^{-|x - y|^2}$, it is

\begin{equation*}
\begin{pmatrix}
1 & e^{-1} & e^{-4} & e^{-9} \\
e^{-1} & 1 & e^{-1} & e^{-4} \\
e^{-4} & e^{-1} & 1 & e^{-1} \\
e^{-9} & e^{-4} & e^{-1} & 1
\end{pmatrix}
\approx
\begin{pmatrix}
1 & e^{-1} & 0 & 0 \\
e^{-1} & 1 & e^{-1} & 0 \\
0 & e^{-1} & 1 & e^{-1} \\
0 & 0 & e^{-1} & 1 \\
\end{pmatrix},
\end{equation*}
where the approximation truncates the kernel so that it contains only three elements and fits in a two qubit register.
This can be block-encoded using the following circuit~\cite[Equation~56]{SCC24}:

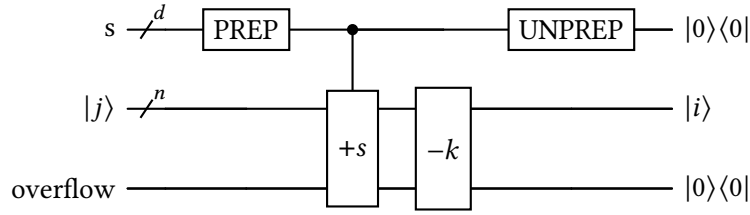
\begin{equation*}
\begin{quantikz}
    \lstick{s} & \qwbundle{d} & \gate{\text{PREP}} & \ctrl{1} & \qw & \gate{\text{UNPREP}} & \qw \rstick{\ket{0}\bra{0}} \\
    \lstick{\ket{j}} & \qwbundle{n} & \qw & \gate[2]{+s} & \gate[2]{-k} & \qw & \qw \rstick{\ket{i}} \\
    \lstick{\text{overflow}} & \qw & \qw & \qw & \qw & \qw & \qw \rstick{\ket{0}\bra{0}} \\
\end{quantikz}
\end{equation*}

Essentially, the \texttt{PREP} and \texttt{UNPREP} blocks load the kernel, while the addition block applies it to the input state.
The subtraction block then adds an offset so that it is centered.
However, while this circuit is conceptually very simple, decomposing it into elementary gates is a lengthy process~\cite{HK25}.
Using \Unitaria{}, this is streamlined significantly, as each of these four blocks is simple to define; though care needs to be taken regarding the different registers on which these operate.
This is solved by taking operations that work only on a subset of the registers and creating the tensor product with the identity for the remaining registers.

First, we create a Numpy array containing the kernel and append a zero to pad the length to a power of two.
Then the \texttt{PREP} and \texttt{UNPREP} blocks are constructed as \texttt{ConstantVector} nodes that load the square root of this padded kernel.
The controlled $+s$ addition block is implemented by the \texttt{IntegerAddition} node, followed by a \texttt{ConstantIntegerAddition} node which adds the constant $-k = -3$ to center the kernel over its third element.
At the end, a slicing operation is used to limit the node to states where the overflow qubit starts and ends in the $\ket{0}$ state.

Combined, this is implemented by the following code:

\begin{lstlisting}[language=Python]
import numpy as np
import unitaria as ut

kernel = np.exp(-(np.arange(-3, 4) / 4) ** 2)
padded_kernel = np.append(kernel, 0)

prep = (ut.Identity(ut.Subspace.from_dim(16))
        & ut.ConstantVector(np.sqrt(padded_kernel)))
add = ut.IntegerAddition(source_bits=3, target_bits=4)
const_add = (ut.ConstantIntegerAddition(bits=4, constant=-3)
             & ut.Identity(ut.Subspace.from_dim(8)))
unprep = ut.Adjoint(prep)

conv = (unprep @ const_add @ add @ prep)[:8, :8]
\end{lstlisting}

Now the tensor product of multiple such one-dimensional Gaussian convolutions yields a multidimensional convolution.
For a more detailed explanation of this code, see \href{https://tequilahub.github.io/unitaria/tutorials/gaussian_convolution/gaussian_conv.html}{the tutorial notebook}.

\subsection{Further examples}

More example code can be found in the repository in the \texttt{examples} folder, including implementations of the numerical experiments from \cite{DP24,DP25,DP25a}.

\section{Conclusion}

This concludes an overview of \Unitaria{}, a software framework providing abstractions for constructing quantum circuits of block encodings.
This document should serve as a summary of the theoretical background, the design, and some basic usage examples of \Unitaria{}.
For further reading, we recommend the documentation available at \url{https://tequilahub.github.io/unitaria}.

While \Unitaria{} already contains all essential functionality, not all circuits can yet be represented in a convenient and ergonomic way.
Improving usability is the subject of ongoing development.
Another issue is that the resulting circuits are usually less efficient than those optimized by hand.
However, \Unitaria{} was designed with optimization workflows in mind. Such optimizers, as well as the implementation of additional specialized nodes, will be the main focus going forward.
Ultimately, the goal is to bring \Unitaria{} to a point where there is no need to think about the low-level circuit implementation anymore, just as one rarely needs to think about assembly in classical programming nowadays.
Instead, developers should be able to focus on implementing the fundamental logic, while \Unitaria{} takes care of optimizing and compiling to circuit level.

Already, \Unitaria{} allows writing quantum algorithms on a mathematically abstract level, easing entry into this field for researchers from different backgrounds and fostering interdisciplinary cooperation in quantum computing.
At the same time, it ensures that research software written today will be able to run on the quantum hardware of the future.

\bibliographystyle{quantum_adapted}
\bibliography{references}

\end{document}